\documentclass[aps,prb,twocolumn,superscriptaddress]{revtex4}
\usepackage{graphicx}
\newcommand{\yhb}{YB$_{6}$}

\newcommand{\msr}{$\mu$SR}

\begin{document}


\title{Microscopic properties of vortex state in YB$_6$ probed by muon spin rotation}


\author{R.~Kadono}
\affiliation{Institute of Materials Structure Science, High Energy Accelerator Research Organization (KEK), Tsukuba, Ibaraki 305-0801, Japan}
\affiliation{Department of Materials Structure Science, 
The Graduate University for Advanced Studies, Tsukuba, Ibaraki 305-0801, Japan}
\author{S.~Kuroiwa}
\affiliation{Department of Physics, Aoyama-Gakuin University, 
Sagamihara, Kanagawa, 229-8558 Japan}
\author{J.~Akimitsu}
\affiliation{Department of Physics, Aoyama-Gakuin University, 
Sagamihara, Kanagawa, 229-8558 Japan}
\author{A.~Koda}
\affiliation{Institute of Materials Structure Science, High Energy Accelerator Research Organization (KEK), Tsukuba, Ibaraki 305-0801, Japan}
\affiliation{Department of Materials Structure Science, 
The Graduate University for Advanced Studies, Tsukuba, Ibaraki 305-0801, Japan}
\author{K.~Ohishi}
\affiliation{Advanced Science Research Center, Japan Atomic Energy Agency,
Tokai, Ibaraki 319-1195, Japan}
\author{W.~Higemoto}
\affiliation{Advanced Science Research Center, Japan Atomic Energy Agency,
Tokai, Ibaraki 319-1195, Japan}
\author{S.~Otani}
\affiliation{Advanced Materials Laboratory, National Institute for Materials Science, 
Tsukuba, Ibaraki 305-0044, Japan}


\date{\today}

\begin{abstract}

Local magnetic field distribution $B({\bf r})$ in the mixed state of a boride superconductor,
\yhb, is studied by muon spin rotation (\msr). A comparative 
analysis using the modified London model and Ginzburg-Landau (GL) model
indicates that the GL model exhibits better agreement with \msr\ data 
at higher fields, thereby demonstrating the importance of  
reproducing the field profile near the vortex cores when the intervortex
distance becomes closer to the GL coherence length.   The temperature and
field dependence of magnetic penetration depth ($\lambda$) does not
show any hint of nonlocal effect nor of low-lying quasiparticle excitation. This suggests that the strong coupling of electrons to the rattling 
motion of Y ions in the boron cage suggested by bulk measurements 
gives rise to a conventional superconductivity with 
isotropic $s$-wave pairing.
Taking account of the present result, a review is provided for 
probing the anisotropy of superconducting 
order parameters by the slope of $\lambda$ against field
($\eta\propto d\lambda/dH$).

\end{abstract}


\maketitle



\section{Introduction}

The revelation of superconductivity in magnesium diboride (MgB$_2$, 
$T_c\simeq39$ K) 
has stimulated renewed interest to other boride superconductors as an alternative
path to novel superconductors with ever higher $T_c$.\cite{Nagamatsu:01}  
Among those, yttrium hexaboride 
(\yhb) currently holds the position of second highest transition temperature 
($T_c\ge7$ K),\cite{Matthias:68} 
thereby drawing considerable attention on the detailed mechanism of 
superconductivity. In particular, it stands as a focus of investigation on the effect of 
rattling motion of alkaline earth metals contained in a relatively large 
boron cage that gives rise to anharmonic phonons having relatively low energy 
($\simeq10^1$ meV).\cite{Lortz:06}  A similar situation is reported in a variety of novel 
superconductors including metal-doped silicon clathrates\cite{Tse:05} and 
$\beta$-pyrochlores.\cite{Hiroi:05}

While \yhb\ is known as a type II superconductor in which the 
Ginzburg-Landau (GL) parameter $\kappa$ ($\equiv\lambda/\xi$,
with $\lambda$ being the magnetic penetration depth and
$\xi=\xi_{\rm GL}$ the GL coherence length) is greater than unity, 
it is characterized by a  
low upper critical  field [$B_{c2}=\mu_0H_{c2}(0)\simeq0.3$ T] and associated 
long $\xi$ ($\sim\xi_0\simeq30$ nm, the BCS coherence length) that is only a few times
smaller than $\lambda$ ($\simeq130$ nm).   
This means that the 
volume fraction occupied by vortex cores ($\simeq \pi\xi^2 H/\Phi_0$, 
where $H$ is the external magnetic field 
and $\Phi_0$ is the flux quantum) is relatively large in the flux 
line lattice (FLL) state over the entire field range, making \yhb\ an ideal stage for
studying the electronic structure of vortex cores in detail. 
In this situation, it is pointed out that the modified London (m-London) model, 
which is known to be an excellent analytical model to describe the spatial 
distribution of magnetic field $B({\bf r})$ for arbitrary $\kappa$
but {\sl at low magnetic 
induction} ($B/B_{c2}<0.25$),\cite{Brandt:88,Fesenko:93} may not be 
appropriate for the basis to extract
physical parameters out of $B({\bf r})$.\cite{Sonier:00}
One of the primary purposes of the present study is to address this potential 
problem by making a comparative analysis of \msr\ data by both m-London
and GL models, which reveals that the latter provides systematically better 
description of magnetic field profile at higher magnetic induction.
Thus, it is inferred from this study that we have to
resort to the GL model for proper understanding of the FLL state over 
the wider field range. Fortunately, however, in contrast to the case of 
classical type II superconductors such as Nb or V ($\kappa\simeq1$)
where one might need numerical approach to obtain $B({\bf r})$ with 
sufficient precision,\cite{Brandt:97,Laulajainen:06} 
the present result suggests that an analytical solution of the GL model 
obtained by variational method\cite{Clem:75,Yaouanc:97} is sufficient for the system with 
reasonably large $\kappa$.

It must be stressed that the magnetic penetration depth 
controlling the actual shape of $B({\bf r})$ corresponds 
to an effective value of the London penetration depth 
determined by the superfluid density $n_s$,
\begin{equation}
\frac{1}{\lambda^2} = \frac{4\pi e^2}{m^*c^2}n_s 
\end{equation}
(where $m^*$ is the effective mass of the charge carriers),  
so that it may vary according to the change in $n_s$ 
due to quasiparticle excitations (breaking of the Cooper pairs) of various origins.
In particular, it is anticipated that the pair breaking due to the quasiclassical Doppler 
shift of the Fermi momenta around vortices would lead to the low energy quasiparticle 
excitation (the nonlinear effect).\cite{Volovik:93} 
Considering that the number of vortices is proportional to the external field, 
it is predicted that $n_s$ is more reduced at higher $H$ due to the Doppler shift.
Another important consequence of the anisotropic order parameter 
is the nonlocal effect due to the variable coherence length [$\xi_0\equiv\xi_0({\bf k})$] 
that may exceed the London penetration depth 
($\lambda_{\rm L}$) over a certain region of the Fermi surface, 
as it is inversely proportional to the 
order parameter [$\xi_0({\bf k})\propto 
\hbar v_{\rm F}/\pi\Delta({\bf k})$, with $v_{\rm F}$ being
the Fermi velocity].\cite{Kosztin:97}
In this situation, the flow of supercurrent is strongly modified 
over the region $\lambda_{\rm L}<\xi_0({\bf k})$, leading to
the change (expansion) of effective $\lambda$.\cite{Amin:98}
Note that this is not limited to the case of nodal gap,
but would be in effect for anisotropic gap 
(or multi-gapped) with the minimum gap 
satisfying $\Delta_{\rm min}<\hbar v_{\rm F}/\pi\lambda_{\rm L}$.
Meanwhile, multi-gapped superconductors have a multitude of 
coherence length and associated vortex core radius within
which the quasiparticles are confined. The presence of 
quasiparticles over the region of larger vortex 
cores would serve as a factor to enhance the effective $\lambda$
more strongly at lower fields, thus mimicking the above mentioned
Volovik effect without gap nodes.

We have shown that the magnetic field dependence of $\lambda$
provides a useful criterion to assess the anisotropy of superconducting 
order parameter; the greater field gradient ($d\lambda/dH$)
corresponds to stronger anisotropy or multi-gapped 
structure in the order parameter.\cite{Kadono:04}
We demonstrate in \yhb\ that $\lambda$ 
deduced from the GL model-based analysis does not depend on the
external magnetic field.  This provides 
strong microscopic evidence for the isotropic superconducting gap in \yhb.

\section{Models for the Field Profile}

In the conventional \msr\ studies of the FLL state, positive muons are implanted
to the superconducting specimen with their initial spin polarization 
perpendicular to the external magnetic field [$\hat{P}(0)\perp\vec{H}$, 
with $\vec{H}\parallel\hat{z}$ in the following],  
so that they may probe $B({\bf r})$ [$\equiv B_z({\bf r})$] 
generated by a periodic array of magnetic vortices which are apart by
$a_0\simeq\sqrt{\Phi_0/H\sin\theta}$ (with the apex angle $\theta=60^\circ$ for
the hexagonal FLL and $=90^\circ$ for the square FLL).
Considering that muon stops at the interstitial sites in the atomic
unit cell of the crystalline lattice which has a periodicity much
shorter than that of vortices 
($a_0\ge89$ nm for $\mu_0H\le0.3$ T assuming a hexagonal FLL), 
one can presume that \msr\ signal in the FLL state 
provides a random sampling of $B({\bf r})$,
\begin{eqnarray}
\hat{P}(t) &=&P_x(t)+iP_y(t) \nonumber\\
 &=&\int_{-\infty}^\infty 
n(B)\exp(i\gamma_\mu Bt-i\phi)dB,\label{Pt}\\
n(B) & = & \langle\delta(B-B({\bf r}))\rangle_{\bf r},
\end{eqnarray}
where $n(B)$ is the spectral density for the internal field defined as a spatial
average ($\langle\:\rangle_{\bf r}$) of the delta function,  
$\gamma_\mu$ is the muon gyromagnetic ratio (= 2$\pi\times 135.53$ MHz/T), and
$\phi$ is the initial phase of rotation.\cite{Brandt:88}
These equations indicate that the real amplitude of the Fourier transformed muon spin 
precession signal, ${\rm Re}\int \hat{P}(t)\exp(-i\gamma_\mu Bt)\gamma_\mu dt$, corresponds to the spectral density, $n(B)$. 

In general, $B({\bf r})$ is obtained by solving the GL equation 
for the field profile and the order parameter $\psi({\bf r})$ in
the self-consistent manner, which usually requires numerical approach.
However, in the case of strong type II superconductors ($\kappa\gg1$), the nonlocal 
character is confined within the immediate vicinity of vortex cores
($|{\bf r}|\le\xi_{\rm GL}$)
and the London approximation can be extended by introducing a cutoff
factor associated with the vanishing $\psi({\bf r})$ at the vortex center.
In the modified London model, $B({\bf r})$ satisfies the equation
\begin{equation}
B({\bf r})-{\rm curl}[\hat{\Lambda}\cdot{\rm curl} B({\bf r})]=\Phi_0\sum_i\rho({\bf r}-{\bf r}_i),
\end{equation}
where $\hat{\Lambda}$ is the tensor related to the penetration
depth and $\rho({\bf r})$ is the source term for the vortices located at the 
position ${\bf r}_i$ in the lattice. 
The solution is given by a sum of the magnetic induction from isolated vortices to yield
\begin{eqnarray}
B({\bf r}) &=& \sum_{\bf K}b({\bf K})\exp(-i{\bf K}\cdot{\bf r})\label{mLondon}\\
b({\bf K}) &=& \frac{B_0}{1+{\bf K}\hat{L}{\bf K}}F(K,\xi_{\rm c})\:,\label{SpFou}
\label{Br}
\end{eqnarray}
where ${\bf K}$ are the vortex reciprocal lattice vectors,
$B_0$ ($\simeq \mu_0H$) is the average internal field, $\hat{L}({\bf K})$ 
is the {\sl effective} London penetration depth, 
$F(K,\xi_{\rm c})$ is the cutoff factor (with $\xi_{\rm c}\propto\xi$ being the cutoff parameter) 
to remove the unphysical divergence in the magnetic field distribution 
near the vortex center: The local London approximation,
$\rho({\bf r})=\delta({\bf r})$, corresponds to $F(K,\xi_c)=1$, where $B({\bf r})$
exhibits a logarithmic divergence. The local approximation also
leads to the reduction of the term ${\bf K}\hat{L}{\bf K}$
into $\lambda^2K^2$ corresponding to the isotropic Fermi surface.
Here, ``local" implies that the electromagnetic 
interaction does not depend on the wave vector (${\bf K}$),
whereas that referring to the source term is primarily meant for $\xi_c=0$.
The form of $F(K,\xi_{\rm c})$ depends on the modelling of 
$\rho({\bf r})$ and associated $\psi({\bf r})$ in the vicinity of vortex cores.
For example, the Gaussian cutoff,
\begin{equation}
F(K,\xi_{\rm c})=\exp\left(-\frac{1}{2}K^2\xi_c^2\right),\label{gausscut}
\end{equation}
which is widely used for \msr\ 
data analysis (valid for $B/B_{c2}<0.25$ and $\kappa\gg1$), 
is derived by the isotropic GL theory
from the Gaussian source term in the London equation\cite{Brandt:88,Brandt:77}
\begin{equation}
\rho({\bf r})=\frac{\Phi_0}{2\pi\xi_0^2}\exp\left(-\frac{r^2}{2\xi_0^2}\right).
\end{equation}
Although this form is presumed to be valid only in the GL limit (i.e., near $T_c$),
it is known that Eq.~(\ref{gausscut}) reproduces the cutoff function 
derived from the most commonly assumed form of the order parameter 
that fits well the solution of the Bogoliubov-de Gennes (BdG) equation,
\begin{equation}
\psi(r)=\psi_{\infty}\tanh\frac{r}{\xi}.
\end{equation}
We also note that a 
renormalzation factor, $\sqrt{1-b}$ (where $b\equiv B/B_{c2}$), was 
introduced to the m-London model in the earlier literatures 
to account for the reduction of the GL order parameter
at higher fields (see below), 
where $\lambda$ and $\xi_c$ in Eq.~(\ref{SpFou}) are divided by 
this factor.\cite{Brandt:88,Sonier:00} However, it turns out throughout 
this work that this renoramlization is too strong to be consistent
with our experimental observation, and therefore disregarded 
in the following analysis.

\begin{figure}[t]
\includegraphics[width=1.0\linewidth]{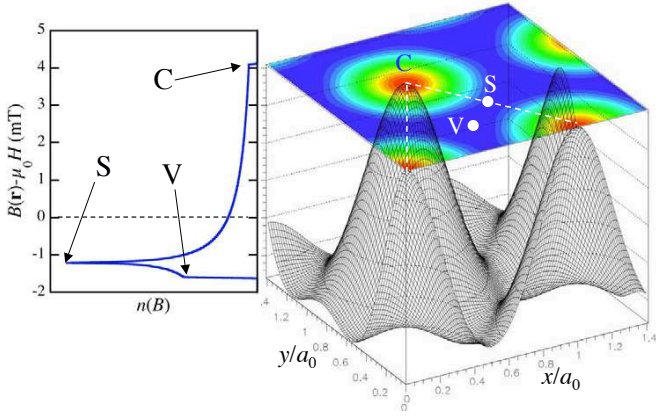}%
\caption{Density distribution function ($n(B)$, left) 
and corresponding spatial field profile ($B({\bf r})$, right) 
for a hexagonal flux line lattice obtained 
from the Ginzburg-Landau model; $\lambda=125$ nm, $\xi=30$ nm,
and $B_0=0.15$ T. The distance between flux lines, $a_0$, is 12.6 nm. The points marked
 as `C', `S', `V' in $n(B)$ respectively correspond to those in 
 $B({\bf r})$. Thus, despite its one-dimensionality, $n(B)$ has
 a site-selective sensitivity in the unit cell of FLL.
\label{nb}}
\end{figure}
\begin{figure}[t]
\includegraphics[width=0.8\linewidth]{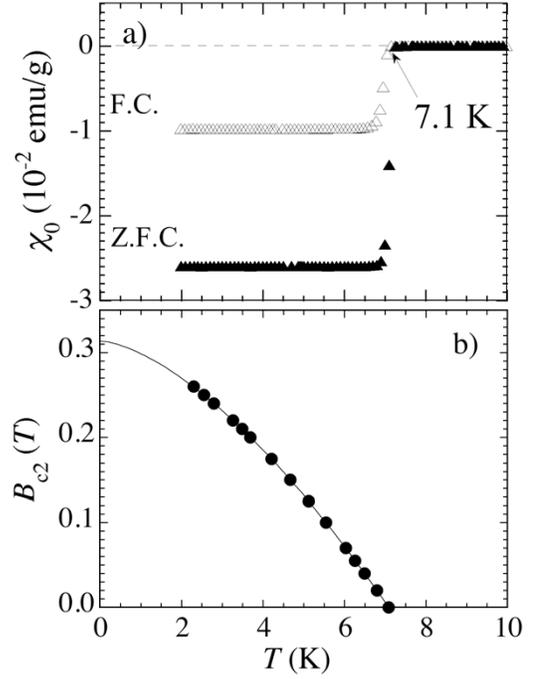}%
\caption{Superconducting transition in \yhb\ observed by 
(a) magnetic susceptibility ($\mu_0H=1$ mT), and (b) the
upper critical field determined by specific heat. 
Solid curve in (b) is fit by a power law (see text).
\label{bulk}}
\end{figure}

On the other hand,
an approach using variational method has been made to obtain the
analytical GL solution for $B({\bf r})$ by assuming
\begin{equation}
\psi(r)=\psi_{\infty}\frac{r}{\sqrt{r^2+\xi_v^2}},\label{psiGL}
\end{equation}
which leads to a squared Lorentzian source term\cite{Amin:98}
\begin{equation}
\rho({\bf r})=\frac{\Phi_0}{\pi}\frac{\xi_v^2}{(r^2+\xi_v^2)^2},
\end{equation} 
and associated cutoff function (for $B\ll B_{c2}$)
\begin{equation}
F(K,\xi_v)=uK_1(u),\:\: u=\sqrt{K^2\xi_v^2+\xi_v^2/\lambda^2},
\end{equation}
where $K_1(x)$ is the modified Bessel function and $\xi_v\simeq\sqrt{2}\xi_0$
is the variational parameter.\cite{Clem:75}  This model was extended to 
higher fields and to anisotropic order parameters by Hao {\it et al}.\cite{Hao:91}
to consider the reduction of order parameter due to the
overlap of vortex cores, yielding the Fourier component
\begin{equation}
b({\bf K}) = B_0\frac{vK_1(u)}
{uK_1(v)},\:\: 
u=\sqrt{K^2\xi_v^2+v^2},\:
v=\frac{\xi_v}{\lambda}f_\infty,\label{SpFou2}
\end{equation}
where $\xi_v$ and $f_\infty$ are the variational parameters with the latter
representing the field-dependent order parameter ($\psi_\infty\rightarrow
f_\infty\psi_\infty$; $f_\infty\rightarrow1$ for the dilute vortex limit).
Under a condition of $\lambda^2K^2_{\rm min}\gg1$, where ${\bf K}_{\rm min}$
is the smallest nonzero reciplocal lattice vector,
Eq.~(\ref{SpFou2}) is simplified for the extreme type II case 
($\kappa\gg1$) using an approximation $K_1(x)\simeq1/x$ to yield\cite{Yaouanc:97}
\begin{equation}
b({\bf K}) \simeq B_0f_\infty^2\frac{uK_1(u)}{\lambda^2K^2},\label{bessel}
\end{equation}
where
\begin{eqnarray}
f_\infty^2 &=& 1-b^4,\:\:b\equiv B/B_{c2}\\
u^2 &=& K^2\xi_v^2 \simeq 2K^2\xi^2(1+b^4)[1-2b(1-b)^2].\label{corerad}
\end{eqnarray}
In the present case of \yhb, $K_{\rm min}$ for a hexagonal
FLL is $4\sqrt{3}/a_0\simeq
0.015$ nm$^{-1}$ (where $a_0\simeq455$ nm for the lowest 
field $B_{c1}\simeq0.02$ T), which leads to $\lambda^2K_{\rm min}^2\simeq3.9$ 
($\lambda\simeq130$ nm), indicating that $\lambda^2K_{\rm min}^2$ is considerably 
greater than unity. Moreover, $K_{\rm min}^2\xi^2\simeq0.2$
($\xi\simeq30$ nm), which is again larger than $v^2\le\kappa^{-2}\simeq0.05$
to justify the approximation, $u\simeq K\xi_v$ [see Eq.~(\ref{SpFou2})].
Thus, \yhb\ satisfies the conditions for Eqs.~(\ref{bessel})--(\ref{corerad})
to be employed as a model to describe $B({\bf r})$.

An example of $B({\bf r})$ and corresponding $n(B)$ calculated 
by the GL model is shown in Fig.~\ref{nb}. 
Regardless of the details in the modeling of $B({\bf r})$, 
it predicts an asymmetric field
profile for $n(B)$ characterized by a negatively shifted sharp peak due
to the van Hove singularity associated with the saddle points of $B({\bf r})$
(marked as `S' in Fig.~\ref{nb}),
and an adiabatic tail towards higher fields where the maximal field is determined
by $B(|{\bf r}-{\bf r}_i|\sim\xi)$.
The nearly one-to-one correspondence
between the characteristic points of $n(B)$ and those on $B({\bf r})$ 
makes it feasible to deduce complex physical parameters including 
$\lambda$ and $\xi$ from a single \msr\ spectrum. 
Meanwhile, information on the structure of FLL must be provided
by other techniques (e.g., small angle neutron scattering). 
Since no such information seems to be available for \yhb,
we presume in the following that a hexagonal FLL is realized.

While one might argue that the m-London model should be
replaced by the GL model simply because of the wider field
range of applicability, we point out that the physical 
meaning of $\xi_v$ in the GL model has some uncertainty
in the sense that it has a complicated field 
dependence as a cutoff parameter [see Eq.~(\ref{corerad})],
yielding different values at different fields for a 
common $\xi$.  This casts some ambiguity, particularly
on the behavior of vortex core radius as a function of
external field. The GL model has another disadvantage in the 
practical application that it requires significant computing
resources (CPU power, in particular) for the precise treatment.
The m-London model, in contrast, is relatively free from 
such ambiguity of model parameters, and it can be implemented 
with ordinary computing environment. Thus, the m-London 
model would remain to be a useful basis for the application
of \msr\ to the study of the mixed state, as far as precaution 
is taken on the limit of its validity.

Finally, it may be worthy of mention that $F(K,\xi_v)$ would be
also dependent on the details of the Fermi surface, and  
thereby it may be subject to further correction
to include ${\bf K}$ dependence.\cite{Nishimori:04}
While the effect of such anisotropy within the vortex cores may 
be small at the dilute vortex limit, it may be important 
at fields close to $B_{c2}$ in anisotropic superconductors. 

\begin{figure}[t]
\includegraphics[width=1.0\linewidth]{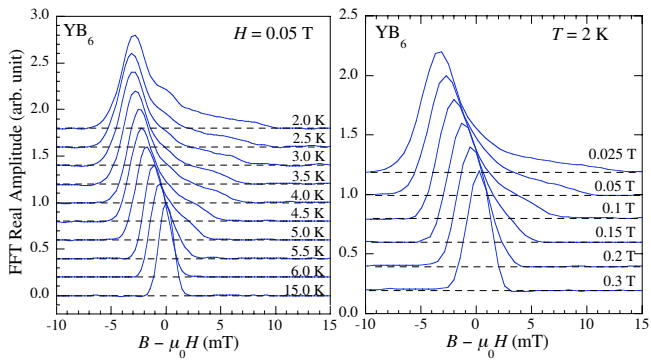}%
\caption{Real amplitude of the fast Fourier transform (FFT) of \msr\ time spectra
under several temperature/magnetic fields ($\mu_0H$) in the mixed state of \yhb\  
(corresponds to $n(B)$).
The small bump at $B-\mu_0H=0$ seen in the spectra at low temperature/field
is due to background signal from muons stopped in the sample holder. 
\label{fftall}}
\end{figure}

\section{Experimental Details}

Single crystals of \yhb\ were prepared by the floating zone
technique, where the details of the crystal growth is found 
elsewhere.\cite{Otani:00} 
Magnetization, specific heat, 
and electronic resistivity were measured prior to \msr\ experiment to 
evaluate the bulk properties of the specimen. 
As shown in Fig.~\ref{bulk}, the result indicates that the crystal has 
a superconducting transition temperature of $T_c=6.95(15)$ K (defined
at the midpoint of change in $\chi_0$) and an upper critical field of
$B_{c2}(0)=0.31$ T.
Other physical properties 
deduced from those measurements are
summarized in Table \ref{tab1}. 
The ratio $2\Delta(0)/k_BT_c\simeq3.67$ 
(deduced from the jump of the electronic specific heat at $T_c$) is close to
the BCS value of 3.53. The mean-free path is estimated to be 
$\sim$2.83 nm from the
residual resistivity ratio. This is much shorter than 
$\xi_0$ [$=(\Phi_0/2\pi B_{c2}(0))^{1/2}$],
thereby indicating that the superconductivity is in the dirty
limit. This strongly suggests that nonlocal effect in terms of
${\bf K}$ dependence, if it exists at all, must be masked 
by the impurity scattering in this specimen.

\begin{table}[b]
\begin{tabular}{cccccccc}
\hline\hline
 $T_c$ (K) & $\frac{2\Delta(0)}{k_BT_c}$ & rrr & $B_{c1}$ (mT) & $B_{c2}$ (T) 
 & $l$ (nm) & $\lambda_{\rm L}$ (nm) & $\xi_0$ (nm) \\
 \hline
 6.95(15) & 3.67 & 4.62 & 18.5 & 0.31
 & 2.83 & 134(2) & 33(1) \\
 \hline\hline
\end{tabular}
\caption{Bulk properties of \yhb\ used for the \msr\ experiment,
where $T_c$ is the superconducting transition temperature (defined 
by the midpoint of $\chi_0$), $\Delta(0)$ is the energy gap extrapolated
to zero temperature, rrr is the
residual resistivity ratio, $B_{c1}$ and $B_{c2}$ are the lower
and upper critical field, $l$ is the mean free path (deduced from rrr), 
$\lambda_{\rm L}$ is the London penetration depth (from $B_{c1}$), and
$\xi_0$ is the BCS coherence length 
(from $B_{c2}$).\label{tab1}}
\end{table}

Conventional $\mu$SR measurements under a transverse field
(TF) were carried out on the
M15 beamline at the Tri-University Meson Facility (TRIUMF, Vancouver, Canada), 
where the field direction was perpendicular to the crystal
plane (normal to [001] axis) covering 
an area of 8 mm $\times$ 8 mm on the sample holder ($H\parallel\hat{z}$ and [001]). 
A positive muon beam with the initial polarization normal to 
the $\hat{z}$ axis was implanted into the specimen,
and the time evolution of muon spin polarization [$\hat{P}(t)$] was
monitored by detecting the decay positrons emitted preferentially towards
the direction of $\hat{P}(t)$. Each measurement was done by 
accumulating $2\sim4\times10^7$ positron events on  
two pairs of scintillation counters (for $\hat{x}$ and $\hat{y}$ directions) 
that yield \msr\ signals
\begin{equation}
\hat{A}(t)=A_x(t)+iA_y(t)=A_0\hat{P}(t),
\end{equation}
where $A_x(t)$ and $A_y(t)$ are the decay positron asymmetry for the 
respective pair of counters with $A_0$ being the total asymmetry.
Measurements on the vortex state were made under field-cooled conditions
to minimize the effect of flux pinning that may give rise to additional
broadening of the \msr\ lineshape.

\begin{figure}[t]
\includegraphics[width=1.0\linewidth]{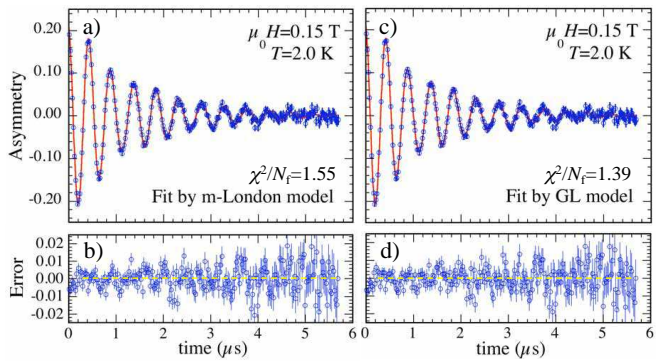}%
\caption{Time revolution of $\mu$-e decay asymmetry [$A_x(t)$, circles] in 
\yhb\ at $\mu_0H=0.15$ T and T=2.0 K displayed on a rotating reference
frame of 18 MHz, where solid curves in a) and c) are
fits assuming $B({\bf r})$ obtained by the m-London model
($\lambda=153$ nm, $\xi_c=25.6$ nm) and GL model ($\lambda=123$ nm, $\xi=29.6$ nm), respectively. 
Fit errors are shown in b) and d) as difference patterns between data and fit. 
\label{tspec}}
\end{figure}

\section{Result}
\subsection{Comparison of models: Field dependence of $\lambda$ and $\xi$}\label{sectIV}
\begin{figure}[t]
\includegraphics[width=0.8\linewidth]{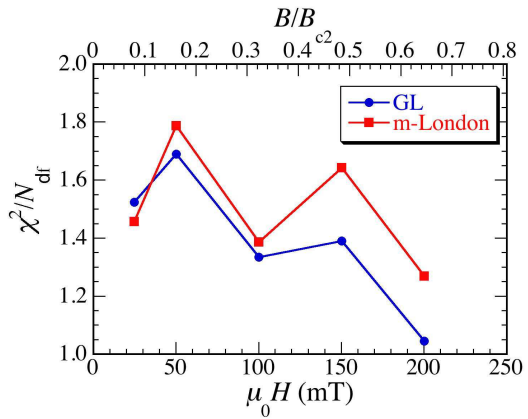}%
\caption{Residual $\chi^2$ divided by the number of degree of freedom ($N_{\rm df}$). 
\label{chi}}
\end{figure}

Examples of the fast Fourier transform (FFT) of \msr\ spectra 
at several external magnetic
fields are shown in Fig.~\ref{fftall}, where one can readily observe the asymemtric
lineshape with slight smearing due to the limited time
window for FFT (0--6 $\mu$s). 
The peaks at lower field represent the van Hove singularity 
due to the saddle points of $B({\bf r})$
(`S' in Fig.~\ref{nb}). 
additional peak at $B-\mu_0H=0$ is 
due to muons which missed the specimen and stopped in the positron counters
surrounding the specimen. The value of  
$\mu_0H$ is experimentally determined by the spectrum
of each field above $T_c$.
Based on the least-square method with appropriate consideration
of the statistical uncertainty, the \msr\ spectra are compared (in time domain) with
those calculated by using Eqs.(\ref{Pt})--(\ref{Br}),
\begin{eqnarray}
\hat{A}(t) &=& A_0\int_{-\infty}^\infty [(1-f_{\rm b})
e^{-(\sigma_{\rm p}^2+\sigma_{\rm n}^2)t^2}
n(B)\:\:\:\nonumber\\
  &  & \:\: + \:\:f_{\rm b}\delta(B-\mu_0H)]e^{i\gamma_\mu Bt-i\phi}dB,
\end{eqnarray}
from which a set of parameters for $n(B)$, $\lambda$, and $\xi_c$ [or $\xi$ 
in Eq.~(\ref{corerad}) for the GL model] is deduced. Here,
$\sigma_{\rm p}$ is the additional 
relaxation due to random flux pinning, 
$\sigma_{\rm n}$ is that due to random local fields from
nuclear magnetic moments,
and $f_{\rm b}$ denotes the fractional yield of the background signal. 
$\sigma_{\rm n}$ ($\simeq0.17$ MHz) 
was determined from the \msr\ spectrum above $T_c$, and
$f_{\rm b}$ was less than 3 \% of the total asymmetry.
The background comes from muons stopping in the 
positron counters (which survived due to the 
finite deficiency of {\sl veto} logic circuits to eliminate
such positron events) and thereby showing 
negligibly small relaxation rate.
We found that the average internal field ($B_0$) showed 
a small positive shift ($\le1$ mT) from the external
field ($\mu_0H$)
with randomly varying magnitude. 
We attribute this to the effect of random pinning of
vortices and associated demagnetization 
which seems to be strong in this
specimen as suggested by relatively large $\sigma_p$ 
($\simeq$0.2--0.5 MHz at any field/temperature).
Thus, we allowed $B_0$ to vary slightly ($\le1$ mT) to 
obtain the best fit.
Fig.~\ref{tspec} shows an example of analysis in the time domain, where 
fits by the m-London and GL models are displayed for 
the data at $\mu_0H=0.15$ T and $T=2.0$ K.  Although both models
yield reasonable fit to data, the ratio of $\chi^2$ to the number of degree
of freedom is considerably reduced from 1.55 for the m-London model
to 1.39 for the GL model: one may observe the scatter in b) is slightly
reduced in d).  As is evident in Fig.~\ref{chi}, this tendency is more
clearly observed at higher fields, and it is consistent with the presumption
that the GL model provides better description over the higher field region. Meanwhile, it should be also stressed that both models
yield almost the same quality of fits at lower fields ($b\le0.3$).

\begin{figure}[t]
\includegraphics[width=1.0\linewidth]{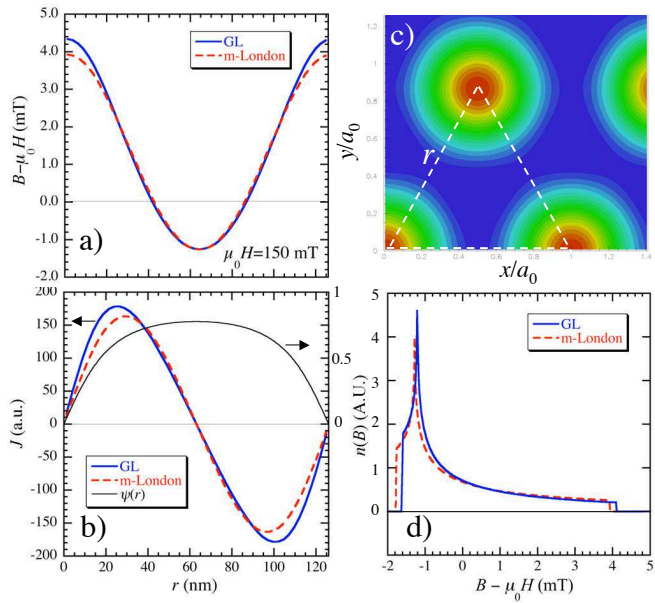}%
\caption{(a) The spatial field profile of $B(r)$ along the straight line
connecting nearest neighboring vortices [shown in (c) by dashed line], 
where $\lambda$ and $\xi_c$ ($\xi$) are taken respectively from the result of 
fits shown in Fig~\ref{tspec}. (b) The spatial profile of the supercurrent 
obtained by the Maxwell's relation [$J(r)={\rm curl}B(r)$]. The normalized 
order parameter [$\overline{\psi}(r)=\psi(r)/\psi(\infty)$] calculated by Eq.~(\ref{psiGL}) 
is also plotted. (d) The density distribution $n(B)$ corresponding to
each curve in (a).
\label{bmap}}
\end{figure}

Figure \ref{bmap} shows comparison between 
those two models for the profiles of $B(r)$, supercurrent calculated by
the Maxwell's relation,
\begin{equation}
J(r)={\rm curl}\{B(r)\}=\frac{dB_z(r)}{dx}-\frac{dB_z(r)}{dy},\label{maxwell}
\end{equation} 
and superconducting order parameter [$\psi(r)/\psi(\infty)$] plotted 
along the straight line connecting the nearest-neighbor vortices, where
the curves in (a)--(b) and (d) 
are calculated by using parameters obtained by fits shown in Fig.~\ref{tspec}. 
The field profile generated
by the GL model keeps increasing towards the vortex
center beyond that by m-London model, yielding smaller effective radius 
($r_0$) for vortex cores defined by the
peak of the supercurrent [$J(r_0)=|J|_{\rm max}$ with $r_0\simeq25$ nm for GL and $\simeq30$
nm for m-London]. Note that the definition of $\xi$ in the GL model is
quite different from $\xi_c$ in the m-London model, and thus they 
cannot be compared directly. Meanwhile, $r_0$ is defined directly from the
field profile using Eq.~(\ref{maxwell}), and thereby it serves as a
common ground for comparison of vortex core size. 
It is noteworthy that, despite relatively small value of $b$ ($\simeq0.5$), 
the order parameter in Fig.~\ref{bmap}(b) 
calculated as a product of Eq.~($\ref{psiGL}$) for two neighboring vortices
exhibits considerable reduction at the center between two vortices.
This again suggests that the GL model must be used for the proper description
of FLL state over the relevant field range.

\begin{figure}[t]
\includegraphics[width=0.65\linewidth]{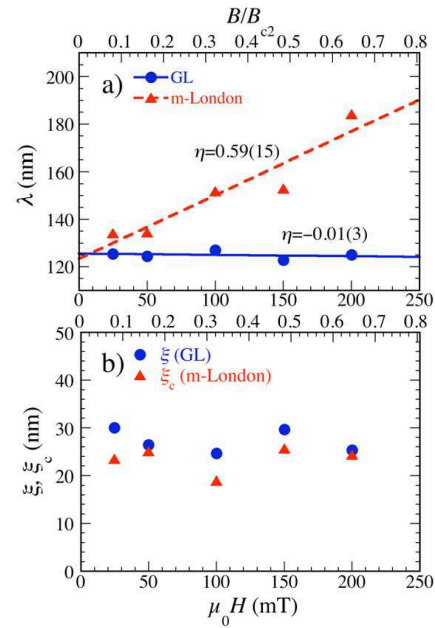}%
\caption{Magnetic field dependence of penetration depth and cutoff parameter 
as determined from the m-London and the GL models. The solid and dashed lines are 
linear fits to the data obtained by the respective models.
\label{lmd-xi}}
\end{figure}

While the difference between two models appears to be localized in
the field profile near the vortex cores, it gives rise to 
change in the deduced value of $\lambda$.  As shown in 
Fig.~\ref{lmd-xi}(a), those obtained by the m-London model 
exhibits a clear tendency of longer $\lambda$ at higher 
fields than m-London model. 
A similar result is reported
for the case of NbSe$_2$ (see Fig.~32 of 
Ref.~[\onlinecite{Sonier:00}]), although the data are 
limited to lower fields ($b<0.31$). 
In the case of \yhb, this leads to qualitative difference
in the dimensionless paramater, $\eta$,
to describe the gradient of $\lambda$ against field, 
which is defined as
\begin{equation}
\lambda(b)=\lambda(0)[1+\eta\cdot b].\label{lmdh}
\end{equation}
Fits of data in Fig.~\ref{lmd-xi}(a) 
using the above relation yields $\eta=0.59(3)$ for 
the m-London model and $\eta=-0.01(3)$ for the GL model.
Since it is anticipated that $\eta\simeq0$ for the case
of isotropic order parameter, the analysis by the GL 
model, which provides better description of $B({\bf r})$,
implies that \yhb\ belongs to the class of conventional BCS 
superconductors with isotropic energy gap.  This in turn 
indicates that a precaution must be taken upon evaluating 
the magnitude of $\eta$ to employ model appropriate 
for the relevant field range of data. We will come back to this
issue later below.  

In contrast to the case of $\lambda$, there
is not much difference in the field dependence
of cutoff parameters between two models, where 
both $\xi_{\rm c}$ and $\xi$ fall into similar
values between 20 and 30 nm. While $\xi$ shows 
a weak tendency of decrease with increasing field,
$\xi_{\rm c}$ does not show any clear trend.  
As mentioned earlier,
the corresponding vortex core radius defined 
as the peak position of supercurrent is slightly 
smaller than $\xi$ in the GL model, while an opposite
tendency is observed for the case of m-London model.
Taking an average of $\xi$ over the observed 
field range [$=27(2)$ nm], we obtain 
the corresponding GL parameter $\kappa=4.7(4)$ in 
good agreement with $\kappa\simeq4.1$ from the bulk property
measurements.
The fact that both $\lambda$ and $\xi$ are independent of $H$ 
means that $\kappa$ is also a constant against $H$.

\subsection{Temperature dependence of $\lambda$ and $\xi$}

\begin{figure}[t]
\includegraphics[width=0.8\linewidth]{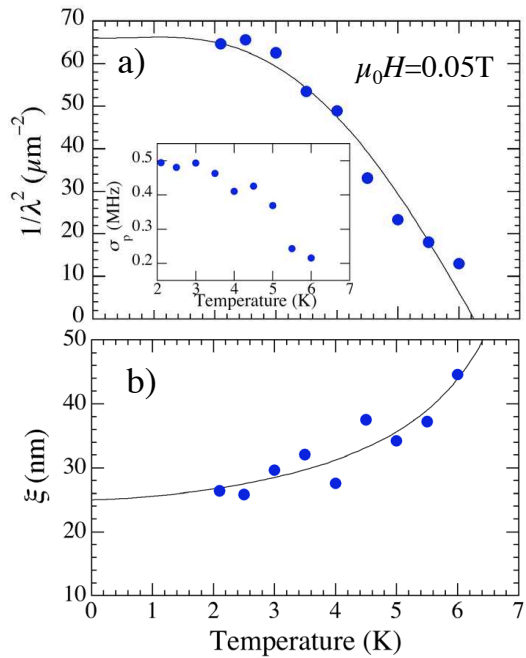}%
\caption{Temperature dependence of $1/\lambda^2$ (a)
and cutoff parameter (b) 
as determined from the GL model. Solid curve in (a) 
is a fit by the weak coupling BCS theory, and that in (b)
is fit by a power law proportional to $[B_{c2}(T)]^{-1/2}$
(see text). Inset: The rate of additional
relaxation due to flux pinning. 
\label{lmdxi-t}}
\end{figure}

Figure \ref{lmdxi-t} shows the temperature dependence of 
$1/\lambda^2$ and $\xi$ determined by the fits of \msr\
spectra at $\mu_0H=0.05$ T, using solutions of the GL model for 
generating $n(B)$.  
Since we showed in the previous section 
that the GL model is superior to m-London 
model, we focus on the result of analysis by the GL 
model in the following.

The solid curve in Fig.~\ref{lmdxi-t}(a) is a fit by the 
weak coupling BCS theory
\begin{equation}
\frac{1}{\lambda^2(T)}=\frac{1}{\lambda^2(0)}
\left[1-2\int_{\Delta}^\infty
\left(-\frac{df}{dE}\right)\frac{EdE}{\sqrt{E^2-\Delta^2}}\right],
\end{equation}
where $f(E)$ is the Fermi distribution function and 
$\Delta=\Delta(T)$ is the temperature-dependent
gap energy.
It reproduces data reasonably well, yielding 
$\lambda(0)=123(2)$ nm and $T_c=6.2(1)$ K. The obtained 
$T_c$ is in excellent agreement with that determined by 
bulk measurements at $\mu_0H=0.05$ T shown in Fig.~\ref{bulk}(b).
The scatter of data points might be attributed to the 
relatively strong pinning that might give rise to a distortion 
of $B({\bf r})$ that is beyond the limit of Gaussian approximation
[i.e., $\exp(-\sigma_{\rm p}^2t^2)$, see the inset of 
Fig.~\ref{lmdxi-t}(a)]. 
Although the absence of data below 2 K does not allow 
definite conclusion, the temperature dependence of 
$1/\lambda^2$ is consistent with the conventional BCS model 
for isotropic $s$-wave pairing, supporting the conclusion
drawn from the field dependence of $\lambda$.
 
The temperature dependence of $\xi$ is displayed in 
Fig.~\ref{lmdxi-t}(b). It exhibits a tendency of 
decrease with decreasing temperature and saturates below 
$T\le4$ K. Similar results are reported for the
cases of NbSe$_2$ (Ref.\onlinecite{Miller:00}) 
and V (Ref.\onlinecite{Laulajainen:06}), where the observed temperature 
dependence has been attributed to the so-called 
Kramer-Pesch effect.\cite{Kramer:74}
In the FLL state, the order parameter
serves as a potential energy [$\propto\psi(r)$, the pair
potential] for the quasiparticles, 
and they are trapped in the vortex cores 
(by the Andreev reflection) to form 
discrete energy levels.
In clean superconductors, it is then predicted 
that the vortex core radius ($\propto\xi$) 
must be dependent on temperature, because the 
higher energy levels, which are occupied by quasiparticles 
at higher temperatures, have greater
extension corresponding to a larger core size.
Unfortunately, however, the experimentally observed
temperature dependence of $\xi$ seems to be much 
weaker than the predicted linear 
dependence, $\xi\sim\xi_0\cdot (T/T_c)$, in those earlier 
examples which are in the clean limit. 
Meanwhile, it is predicted that the effect is 
sensitive to the impurity 
scattering and thereby it would be smeared out in the dirty
limit corresponding to the present case of \yhb. 
This is again in contrast to the observed behavior of 
$\xi$ for $T\ge4$ K, suggesting that it cannot be 
attributed to the Kramer-Pesch effect.

Here, we recall a much simpler 
argument that the weak temperature dependence of
$\xi$ can be naturally explained by considering the 
fact that $B_{c2}$ varies with temperature.\cite{Kadono:01}
Since the GL coherence length near $T_c$ 
is given by the relation
\begin{equation}
\xi_{\rm GL}(T)=\sqrt{\frac{\Phi_0}{2\pi B_{c2}(T)}}.
\label{xi-gl}
\end{equation}
it is natural to assume 
$\xi(T)\simeq\xi_{\rm GL}(T)\propto[B_{c2}(T)]^{-1/2}$.
Thus, the increase of $\xi$ at higher temperatures
might be attributed to the decrease of $B_{c2}(T)$.
A fit of $B_{c2}(T)$ in Fig.~\ref{bulk}(b) by a 
power law, 
\begin{equation}
B_{c2}(T)=B_{c2}(0)[1-\zeta(T/T)^\nu],\label{hc2}
\end{equation}
yields $B_{c2}(0)=0.314(1)$ T, $\zeta=0.970(1)$ and 
$\nu=1.56(1)$
The solid curve in Fig.~\ref{lmdxi-t}(b) is 
the best fit obtained by assuming Eq.~(\ref{xi-gl})
and (\ref{hc2}) with the above parameter values, 
which yields $T_c=7.6(4)$ K. 
The agreement is satisfactory and thereby it 
demonstrates that $\xi$ at 2 K is determined by
$B_{c2}$ near $T_c$.  
This indicates that the thermal fluctuation is 
relatively strong over the relevant temperature 
region ($T/T_c\ge0.29$), and that 
the Kramer-Pesch effect must be searched for 
at temperatures low enough so that one may be 
able to neglect the temperature dependence of 
$B_{c2}(T)$.

\section{Discussion}

\subsection{Superconductivity of \yhb}

Despite the early discovery of superconductivity in \yhb, there 
are not much literatures published so far 
on its superconducting property. Fortunately, a recent 
paper by Lortz {\it et al.}~reports detailed measurements on 
specific heat, resistivity, and thermal expansion.\cite{Lortz:06}
The specific heat in the sperconducting state is that typically 
found for a single-band, isotropic BCS superconductor, which 
we also confirmed on our specimen by similar measurements.
Meanwhile, the electron-phonon interaction turns out to be 
much stronger than that in other boride superconductors 
such as ZrB$_{12}$ and MgB$_2$.  They report an enhanced 
coupling to phonon modes lying near 8 meV in \yhb, which 
is relatively close to $\sim15$ meV of ZrB$_{12}$ (Ref.\onlinecite{Lortz:05}). 
This is to be compared with $\sim60$ meV in MgB$_2$ (Ref.\onlinecite{Geerk:05}), 
thus it provides an explanation for the difference of $T_c$ 
among those three borides. Interestingly, those low-energy
frequency modes are attributed to the vibration of 
Y or Zr atoms loosely bound in oversized boron cages
that is now called ``rattling" motion. 
Here, the longer metal to boron bond length in \yhb\ 
leads to a weaker force constant and larger vibrational amplitude
favorable for higher $T_c$. The reduction of $T_c$ 
under high pressure predicted from the thermal expansion 
measurement, which is also anticipated from the coupling to 
the rattling phonon, has been 
confirmed by the recent experiment on the pressure 
effect.\cite{Khasanov:06}

In the previous section, we have shown that the field 
and temperature dependence of $\lambda$ is fully consistent
with the presence of isotropic order parameter in \yhb.
It does not show any hint of nonlocal effect nor of low-lying 
quasiparticle excitation associated with the 
anomaly in the order parameter. This in turn suggests 
that the strong coupling of electrons to the rattling 
phonons does not necessarily lead to anomaly in the superconducting 
order parameter.  The situation is readily understood
by considering that \yhb\ has a highly symmetric 
three-dimensional (3D) band structure, while MgB$_2$ is characterized
by two-dimensional $\sigma$ band and 3D-$\pi$ band which 
lead to multi-gapped superconductivity (see below).
However, it should be noted that the present specimen 
is in the dirty limit where the anisotropic feature
having the length scale longer than the mean-free path
($l=2.83$ nm, corresponding to $\hbar v_{\rm F}/\pi l\sim7$ meV
assuming that $v_{\rm F}\sim10^5$ m/s) 
must be smeared out by scattering of 
electrons.  Therefore, further study with much better
crystal quality may be needed for the definitive 
conclusion.

\subsection{$\eta$ as a criterion of anomaly in the order parameter}

\begin{figure}[t]
\includegraphics[width=0.85\linewidth]{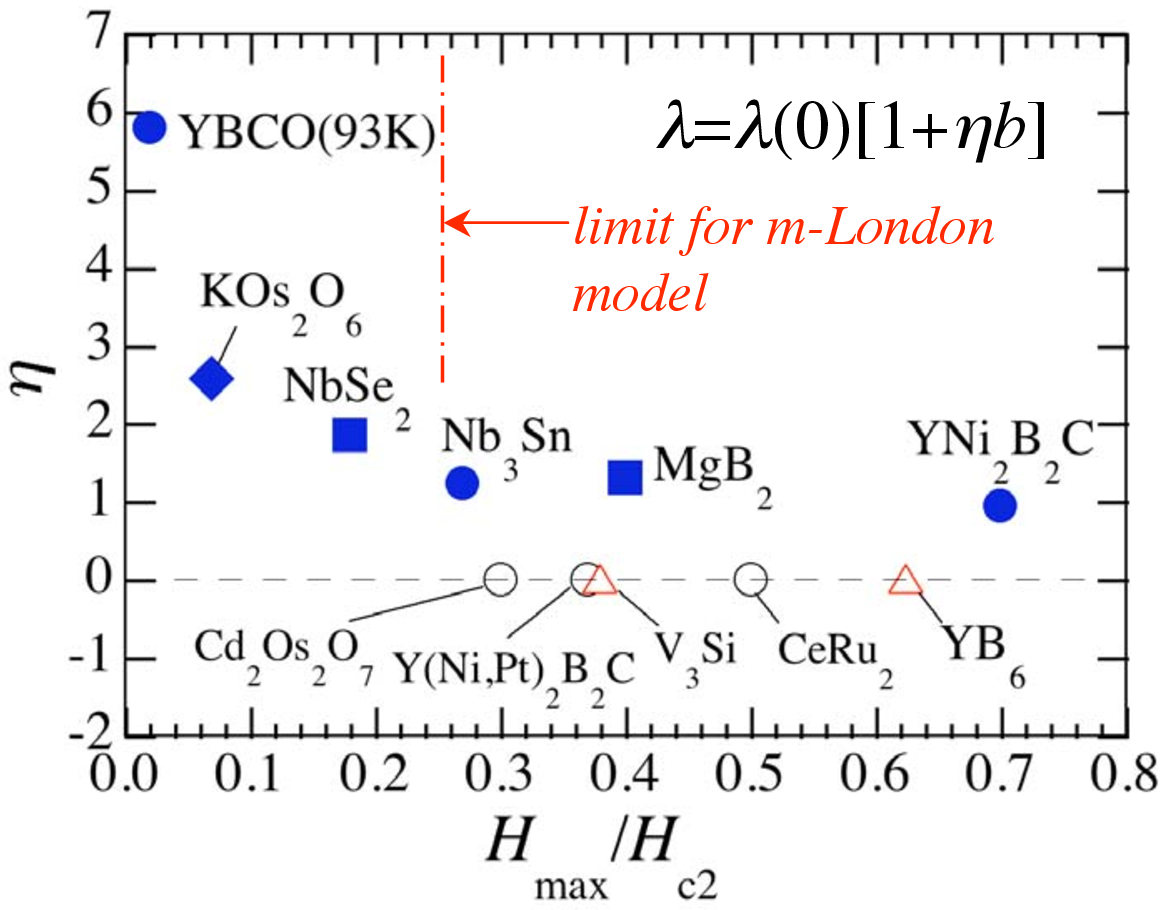}%
\caption{Slope of field dependent $\lambda$ 
plotted against the maximal field of measurement 
normalized by the upper critical field, where 
the m-London model was used to generate $B({\rm r})$
(the relaxation was approximated by a Gaussian damping
for Cd$_2$Os$_2$O$_7$ and MgB$_2$) 
[(Refs.\onlinecite{Sonier:00,Kadono:01,Ohishi:02,Ohishi:03,Ohishi:03b,Kadono:02,Sonier:04,Koda:05,Koda:07,Kadono:06}].
Filled symbols: superconductors with anisotropic gap (circles)
or multigapped (squares).
Open circles and triangles: those with isotropic
gap (where the GL model was used for the latter case). 
\label{eta-hc2}}
\end{figure}

We saw in the section \ref{sectIV} that the slope $\eta$ of the 
field dependence of $\lambda$ obtained by m-London model
may be slightly overestimated when the model is applied beyond the
presumed limit, $B/B_{c2}\le0.25$. 
This may bring about a concern on the past literatures 
discussing the degree of gap anisotropy by the magnitude of $\eta$. 
To make the situation clear, we collected results 
reporting the value of $\eta$ obtained by using m-London model, 
which are shown in Fig.~\ref{eta-hc2}
as a plot of $\eta$ versus the maximal field of measurement
($\mu_0H_{\rm max}$) normalized by $\mu_0H_{c2}$ at the measured
temperature. As a comparison, those obtained by the GL model
is also plotted for V$_3$Si (Ref.\onlinecite{Sonier:04})
and \yhb\ (the present result). 

It is relatively well established by other experimental techniques
that the group of compounds shown by the filled symbols 
in Fig.~\ref{eta-hc2} (those with $\eta\ge1$) have anisotropic superconducting gap 
or multi-gapped, except for the case of KOs$_2$O$_6$ 
(Refs.\onlinecite{Koda:05,Koda:07}) in which the situation is
not clear at this stage. While the observation that $\eta\ge1$
for those compounds supports the use of $\eta$ as a
criterion for the gap anisotropy, the cases of 
YNi$_2$B$_2$C (Ref.\onlinecite{Ohishi:02}) and 
MgB$_2$ (Ref.\onlinecite{Ohishi:03}) might require  
closer examination because of large $H_{\rm max}/H_{c2}$. 
Fortunately for the case of YNi$_2$B$_2$C, 
data with small field step
exist at lower fields ($b\le0.2$), and one can clearly 
observe that $\eta$ remains to be $\simeq1$ within 
the limit of m-London model ($b\simeq0.25$). 
This may be partly attributed to the presence of point 
nodes in the superconducting 
order parameter\cite{Izawa:02,Watanabe:04} that 
would give rise to the nonlocal effect 
in YNi$_2$B$_2$C at higher fields.
Since the nonlocal effect directly affects the 
behavior of $\lambda$, it may be less sensitive to the
difference in the modeling of vortex cores between 
the m-London model and GL model. In particular,
the situation might be that the magnitude of $\eta$ at 
higher fields, which is predominantly 
determined by the nonlocal effect, coincides
with that due to the Volovik effect observed at lower fields. 
Interestingly in YBa$_2$Cu$_3$O$_{6.95}$ (YBCO, having line nodes), 
it is reported that $\eta\simeq2$ for $b\ge0.04$
which is much smaller than $\eta\ge5$ for $b<0.02$ (Ref.\onlinecite{Sonier:00}),
where the magnitude of $\eta$ at higher fields is explained by considering 
the nonlocal effect associated with the line nodes.\cite{Amin:98}
A similar situation would be anticipated with less 
magnitude of $\eta$ at higher fields 
for the case of YNi$_2$B$_2$C with point nodes. The point node also provides 
a natural explanation for small $\eta$ 
compared with that of YBCO at lower fields.

Here, it might be stressed again that there are two different
sources of nonlocal effect, the one associated with 
the anisotropy in the superconducting order parameters
and that associated with the anisotropy in the Fermi surface. 
In general, the latter effect manifests itself as 
the transformation of FLL from hexagonal 
to squared with increasing external 
field, where transformation occurs at relatively low
fields (e.g., $b<0.06$ in YNi$_2$B$_2$C). Meanwhile,
the anisotropy in the order parameter has relatively smaller
energy scales and thereby it emerges at higher fields.
While both yield a qualitatively similar modification 
to the effective London penetration depth ($\hat{L}$), 
their consequences are independent (even competing)
with each other.\cite{Nakai:02}

It is now well established that MgB$_2$ is characterized by
two different energy gaps corresponding to its band structure
($\Delta_{\sigma}\simeq6.8$ meV for $\sigma$ band 
and $\Delta_{\pi}\simeq2.2$ meV for $\pi$ band).\cite{Tsuda:05} 
The analysis in Ref.\onlinecite{Ohishi:03} was made before 
the revelation of the double gap, using the Gaussian approximation 
instead of m-London model. However, it captures anomalous
behavior of $1/\lambda^2$ (proportional to the Gaussian linewidth)
unexpected for the conventional superconductors with single 
gap. 

The Gaussian model is based on a crude 
approximation that the Gaussian damping rate, $\sigma$, 
is primarily determined by the penetration depth
(valid for extreme type II superconductors),
\begin{eqnarray} 
\sigma&=&\
\gamma_\mu{\langle \sum_{\bf K} b({\bf K})^2\rangle}^{1/2}
\propto G(b)\lambda^{-2},\label{sigma}\\
& &G(b)\simeq(1-b)[1+3.9(1-b)^2]^{1/2},\nonumber
\end{eqnarray}
where the function $G(b)$ represents the reduction of 
$\sigma$ mainly due to the overlap of vortices
(proportional to $1-b$) and additional narrowing
due to the contribution of vortex cores.\cite{Brandt:88} 
The presence of two gaps means that there are two 
different upper critical fields corresponding to respective 
coherence length,
\begin{equation}
B_{c2(i)}=\frac{\Phi_0}{2\pi\xi_i}, \:\:(i=\sigma,\pi),
\end{equation}
from which one may anticipate that
\begin{equation}
\sigma\propto [a_\sigma G(b_\sigma)+(1-a_\sigma)G(b_\pi)]\lambda^{-2},
\label{sigma2}
\end{equation}
where $b_i=\mu_0H/B_{c2(i)}$ (with $B_{c2(\pi)}<B_{c2(\sigma)}$), 
$a_\sigma$ is the fractional
weight of $\sigma$ component, and $G(b_\pi)$ must be
set to zero for $b_\pi>1$. The second term in the
above equation, which exhibits steeper field dependence than
the first term, strongly reduces $\sigma$ with increasing 
field over the field range of $0\le \mu_0H <B_{c2(\pi)}$.
From the viewpoint of single $B_{c2}$, such behavior appears
as an anomalous deviation of $\lambda$ from Eq.(\ref{sigma})
and thereby easily identified as an increase of $\eta$ as
reported in Ref.\onlinecite{Ohishi:03}. 
As a matter of fact, recent 
attempt to analyze $\sigma$ assuming a model similar to Eq.~(\ref{sigma2})
reports $\xi_\sigma=5.1(2)$ nm and $\xi_\pi=23(1)$ nm.\cite{Serventi:04}
Considering that the magnitude of $\xi_\pi$ is as large as nearly a half 
of $\lambda$ [=49(4) nm], the enhancement of $\lambda$ 
is physically understood as that due to the 
QP's over a radial region $\xi_\sigma<r<\xi_\pi$ 
around vortex cores. 
This is presumed to be a common situation for multi-gapped
superconductors including another example of 
NbSe$_2$.\cite{Callaghan:05}
We also point out that the smaller $\xi(T)$ (e.g., $\xi_\pi(T)$ 
in MgB$_2$) may exceed $\lambda$ at a certain temperature,
as it increases with $T$ [see Eq.~(\ref{xi-gl}) and Fig~\ref{lmdxi-t}(b)].
Then, there is a possibility that the nonlocal effect may 
set in to further enhance the effective $\lambda$.

Now it is clear that the magnitude of $\eta$ indeed serves as 
a criterion for the anisotropy of superconducting order parameter,
where that obtained from the analysis using the m-London model 
remains useful (particularly those for which the maximal field 
range is within the presumed limit of validity, $b<0.25$).
In this context, the previous result for KOs$_2$O$_6$ may be 
regarded as a strong case for the presence of anisotropic order 
parameter (including multi-gap) in this compound. 
Considering results of other experimental techniques that 
the order parameter is nodeless,\cite{Kasahara:06} 
further analysis of \msr\ data 
by a model based on the double-gap scenario 
is now in progress.

\section{Summary and Conclusion}

We showed that the effective London penetration depth
in \yhb, obtained by analyzing \msr\ data using the
field profile generated by the analytical GL model, 
is independent of external field ($\eta\simeq0$) over a region 
$0<\mu_0H/B_{c2}<0.65$. We made a comparative analysis
using the modified London model and GL model, from which
it was inferred that the GL model should be used 
for the precise determination of $\eta$ when the field
range of measurements extends over the presumed limit of 
validity for the m-London model ($\mu_0H/B_{c2}>0.25$). We also showed
that the temperature dependence of $1/\lambda^2$ is 
perfectly in line with that predicted for
isotropic BCS superconductors.  This, together with the 
observation that $\eta\simeq0$, supports for the 
conclusion drawn from bulk measurements that \yhb\ 
belongs to the class of conventional BCS superconductors
with isotropic $s$-wave pairing.
Following this result, we re-exmined the past reports 
on $\eta$ obtained by the analysis based on the m-London
model (and those based on the Gaussian approximation),
and confirmed that the conclusions drawn for respective
compounds remain unchanged. Thus, the m-London model 
continues to be a useful ground for the \msr\ study of 
vortex state within the limit of its validity.

\section*{Acknowledgements}
We would like to thank the staff of TRIUMF for their technical
support during the \msr\ experiment.
This work was partially supported by a
Grant-in-Aid for Scientific Research on Priority Areas and
a Grant-in-Aid for Creative Scientific Research from the Ministry of
Education, Culture, Sports, Science and Technology of Japan.



\begin{thebibliography}{}

\bibitem{Nagamatsu:01} J. Nagamatsu, N. Nakagawa, T. Muranaka, Y. Zenitani and 
J. Akimitsu: Nature {\bf 410}, 63 (2001).

\bibitem{Matthias:68} B. T. Matthias,  T. H. Geballe, K. Andres, 
E. Corenzwit, G. W. Hull, and J. P. Maita, Science {\bf 159}, 530 (1968).

\bibitem{Lortz:06} R. Lortz, Y. Wang, U. Tutsch, S. Abe, C. Meingast, P. Popovich, W. Knafo, 
N. Shitsevalova, Yu. B. Paderno, and A. Junod, Phys. Rev. B {\bf 73}, 024512 (2006).

\bibitem{Tse:05} J. S. Tse, T. Iitaka, T. Kume, H. Shimizu, K. Parlinski,
H. Fukuoka, and S. Yamanaka, Phys. Rev. B {\bf 72}, 155441 (2005).

\bibitem{Hiroi:05} Z. Hiroi, S. Yonezawa, T. Muramatsu, J. Yamaura, and Y. Muraoka,
J. Phys. Soc. Jpn. {\bf 74}, 1255 (2005).

\bibitem{Brandt:88} E. H. Brandt, 
Phys. Rev. B {\bf 37}, R2349 (1988).

\bibitem{Fesenko:93} V. Fesenko, V. Gorbunov, A. Sidorenko, 
and V. Smilga, 
Physica C {\bf 211}, 343 (1993).

\bibitem{Sonier:00} J. E. Sonier, J. H. Brewer, and R. F. Kiefl,
Rev. Mod. Phys. {\bf 72},  769 (2000).

\bibitem{Brandt:97} E. H. Brandt, 
Phys. Rev. Lett. {\bf 78}, 2208 (1997).

\bibitem{Laulajainen:06} M. Laulajainen, F. D. Callaghan, 
C.V. Kaiser, and J. E. Sonier, 
Phys. Rev. B {\bf 74}, 054511 (2006). This paper seems to have some
problem in their data analysis, as it reports 
$\lambda$ and $\xi$ obtained by comparative analysis 
using m-London model and two GL models 
(analytical and iterative) which are very close with each other, 
while they demonstrate poor agreement of 
$B(r)$ between those three models for a common set of 
$\lambda$ and $\xi$.

\bibitem{Clem:75} J. R. Clem, J. Low Temp. Phys. {\bf 18}, 427 (1975).

\bibitem{Yaouanc:97} A. Yaouanc, P. Dalmas de R\'eotier, and E. H. Brandt,
Phys. Rev. B {\bf 55}, 11107 (1997).

\bibitem{Volovik:93} G. E. Volovik, 
Pis'ma Zh. Eksp. Teor. Fiz. {\bf 58}, 457 (1993)
 [{\it JETP Lett.} {\bf 58}, 469 (1993)].
 
\bibitem{Kosztin:97} I. Kosztin and A. J. Leggett, 
Phys. Rev. Lett. {\bf 79}, 135 (1997).

\bibitem{Amin:98} M.~H.~S. Amin, I. Affleck, and M. Franz,
Phys. Rev. B {\bf 58}, 5848 (1998).

\bibitem{Kadono:04} R. Kadono,
J. Phys.: Condens. Matt. {\bf 16}, S4421 (2004).

\bibitem{Brandt:77} E. H. Brandt, J. Low Temp. Phys. {\bf 24}, 709 (1977). 

\bibitem{Hao:91} Z. Hao, J. R. Clem, M. W. McElfresh, L. Civale, 
A. P. Malozemoff, and F. Holtzberg, 
Phys. Rev. B {\bf 43}, 2844 (1991).

\bibitem{Nishimori:04} H. Nishimori, K. Uchiyama, S. Kaneko, 
A. Tokura, H. Takeya1, K. Hirata and N. Nishida, 
J. Phys. Soc. Jpn. {\bf 73}, 3247 (2004).

\bibitem{Otani:00} S. Otani,
M. M. Korsukova, T. Mitsuhashi, and N. Kieda,
J. Crystal Growth {\bf 217}, 378 (2000).

\bibitem{Miller:00} R. I. Miller, R. F. Kiefl, J. H. Brewer,
J. Chakhalian, S. Dunsiger, G. D. Morris, J. E. Sonier,
and W. A. MacFarlane, Phys. Rev. Lett. {\bf 85}, 1540 (2000).

\bibitem{Kramer:74} L. Kramer and W. Pesch, 
Z. Phys. {\bf 269}, 59 (1974).

\bibitem{Kadono:01} R. Kadono, W. Higemoto, A. Koda, K. Ohishi, T. Yokoo, J. Akimitsu, M. Hedo, Y. Inada, Y. Onuki, 
E. Yamamoto, and Y. Haga, 
Phys. Rev. B {\bf 63}, 224520 (2001).

\bibitem{Lortz:05} R. Lortz, Y. Wang, S. Abe, C. Meingast, Y. B. Paderno, 
V. Filippov, and A. Junod, Phys. Rev. B {\bf 72}, 024547 (2005).

\bibitem{Geerk:05} J. Geerk, R. Schneider, G. Linker, A. G. Zaitsev, R. Heid, K.-P. Bohnen, and H. v. L\"ohneysen,
Phys. Rev. Lett. {\bf 94}, 227005 (2005).

\bibitem{Khasanov:06} R. Khasanov, P. S. H\"afliger, N. Shitsevalova, A. Dukhnenko, R. Br\"utsch, and H. Keller,
Phys. Rev. Lett. {\bf 97}, 157002 (2006).

\bibitem{Sonier:04} J. E. Sonier, F. D. Callaghan, R. I. Miller, 
E. Boaknin, L. Taillefer, R. F. Kiefl, J. H. Brewer, K. F. Poon,
and J. D. Brewer, 
Phys. Rev. Lett. {\bf 93}, 017002 (2004).

\bibitem{Ohishi:02} K. Ohishi, K. Kakuta, J. Akimitsu, W. Higemoto, R. Kadono, J. E. Sonier, A. N. Price, R. I. Miller, R. F. Kiefl, M. Nohara, H. Suzuki and H. Takagi,  
Phys. Rev. B {\bf 65}, 140505(R) (2002).

\bibitem{Ohishi:03} K. Ohishi, T. Muranaka, J. Akimitsu, A. Koda, W. Higemoto, and R. Kadono,
J. Phys. Soc. Jpn. {\bf 72}, 29 (2003).

\bibitem{Ohishi:03b} K. Ohishi, K. Kakuta, J. Akimitsu, A. Koda, W. Higemoto, R. Kadono, J. E. Sonier, A. N. Price, R. I. Miller, R. F. Kiefl, M. Nohara, H. Suzuki and H. Takagi,
Physica B {\bf 326}, 364 (2003).

\bibitem{Kadono:02} R. Kadono, W. Higemoto, A. Koda, Y. Kawasaki, 
M. Hanawa, and Z. Hiroi, 
J. Phys. Soc. Jpn. {\bf 71}, 709 (2002).

\bibitem{Koda:05} A. Koda, W. Higemoto, K. Ohishi, S. R. Saha,
R. Kadono, S. Yonezawa, Y. Muraoka, and Z. Hiroi, 
J. Phys. Soc. Jpn. {\bf 74}, 1678 (2005).

\bibitem{Koda:07} A. Koda, K. H. Satoh, S. Takeshita, R. Kadono, 
K. Ohishi, W. Higemoto, S. R. Saha, Y. Kawasaki, 
T. Minami, S. Yonezawa, Z. Hiroi, and Y. Muraoka, 
unpublished.

\bibitem{Kadono:06} R. Kadono, K. H. Satoh, A. Koda, T. Nagata, H. Kawano-Furukawa, J. Suzuki, M. Matsuda, K. Ohishi, W. Higemoto, S. Kuroiwa, H. Takagiwa, and J. Akimitsu, 
Phys. Rev. B {\bf 74}, 024513 (2006).

\bibitem{Izawa:02} K. Izawa, K. Kamata, Y. Nakajima, Y. Matsuda, 
T.Watanabe, M. Nohara, H. Takagi, P. Thalmeier, and K. Maki,
Phys. Rev. Lett. {\bf 89}, 137006 (2002).

\bibitem{Watanabe:04} T. Watanabe, M. Nohara, T. Hanaguri, 
and H. Takagi, 
Phys. Rev. Lett. {\bf 92}, 147002 (2004).

\bibitem{Nakai:02} See, for example, N. Nakai, P. Miranovi\'c, 
M. Ichioka, and K. Machida, 
Phys. Rev. Lett. {\bf 89}, 237004 (2002).

\bibitem{Tsuda:05} S. Tsuda, T. Yokoya, T. Kiss, T. Shimojima, S. Shin, 
T. Togashi, S. Watanabe, C. Zhang, C. T. Chen, S. Lee, H. Uchiyama, 
S. Tajima, N. Nakai, and K. Machida,
Phys. Rev. B {\bf 72}, 064527 (2005).

\bibitem{Serventi:04} S. Serventi, G. Allodi, R. De Renzi, 
G. Guidi, L. Roman\`o, P. Manfrinetti, A. Palenzona, 
Ch. Niedermayer, A. Amato, and Ch. Baines,
Phys. Rev. Lett. {\bf 93}, 217003 (2004).

\bibitem{Callaghan:05} F. D. Callaghan, M. Laulajainen, 
C. V. Kaiser, and J. E. Sonier, 
Phys. Rev. Lett. {\bf 95}, 197001 (2005). In this paper, the authors
argue that the QP's are delocalized in accordance with the shrinkage
of vortex core radius. They further maintain that the situation is
common to the case of V$_3$Si (see Fig.~4 of this paper). However,
we point out that the field dependence of $\lambda$ in V$_3$Si reported 
in their previous work\cite{Sonier:04} seems  
inconsistent with their interpretation; it remains independent of 
field for $\mu_0H>1.5$ T where the core radius undergoes shrinkage.

\bibitem{Kasahara:06} Y. Kasahara, Y. Shimono, T. Shibauchi, 
Y. Matsuda, S. Yonezawa, Y. Muraoka, and Z. Hiroi,
Phys. Rev. Lett. {\bf 96}, 247004 (2006).


\end{thebibliography}
\end{document}